\documentclass{PoS}

\usepackage{graphicx, fancybox}
\usepackage{amsmath, amssymb, mathrsfs, bm}
\usepackage[T1]{fontenc}
\usepackage[table]{xcolor}
\usepackage{ulem}
\usepackage{cite}

\newcommand{\red}[1]{\textcolor{red}{#1}}
\newcommand{\blue}[1]{\textcolor{blue}{#1}}

\title{
\begin{picture}(0,0)(0,0)%
 \put(220,90){\makebox(0,0)[l]{\textnormal{\normalsize 
 J-PARC-TH-0172, RIKEN-iTHEMS-Report-19}}}%
\end{picture}%
Stress distribution in quark--anti-quark and single quark systems at nonzero temperature
}

\ShortTitle{Stress distribution in quark--anti-quark and single quark systems at nonzero temperature}

\author{\speaker{Ryosuke Yanagihara},$^a$ Takumi Iritani,$^b$
Masakiyo Kitazawa,$^{a,c}$ Masayuki Asakawa,$^a$ Tetsuo Hatsuda$^{d}$
(FlowQCD collaboration)
\\
         \llap{$^a$} Department of Physics, Osaka University, Toyonaka,
	 Osaka 560-0043, Japan
	 \\
	 \llap{$^b$}RIKEN Nishina Center, RIKEN, Wako, 351-0198, Japan
	 \\
	 \llap{$^c$}J-PARC Branch, KEK Theory Center, Institute of Particle
	 and Nuclear Studies, KEK, 203-1, Shirakata, Tokai,
	 Ibaraki 319-1106, Japan
	 \\
	 \llap{$^d$} RIKEN Interdisciplinary Theoretical and Mathematical
	 Sciences Program (iTHEMS), RIKEN, Wako 351-0198, Japan
	 \\
         E-mail: \email{yanagihara@kern.phys.sci.osaka-u.ac.jp}}

\abstract{
We explore the distribution of the energy momentum tensor (EMT) 
around quark--anti-quark and single quark at nonzero temperature 
in SU(3) Yang-Mills gauge theory by extending
our previous study~\cite{Yanagihara:2018qqg} on the EMT distribution 
in static quark--anti-quark systems in vacuum. 
We discuss the disappearance of the flux tube structure observed 
in the vacuum simulation. We investigate the total stress acting 
on the mid-plane between a quark and an anti-quark 
and show that it agrees with the force obtained from the derivative 
of the free energy. The color Debye screening effect 
in the deconfined phase is also discussed in terms of the EMT distribution.
}

\FullConference{37th International Symposium on Lattice Field Theory - Lattice2019\\
		16-22 June 2019\\
		Wuhan, China}

\begin{document}

\section{Introduction}
The energy-momentum tensor (EMT) 
$\mathcal{T}_{\mu\nu}(x)$
plays crucial roles in various fields in physics including 
gravitational theory, hydrodynamics, and elastic body.
Among the components of EMT, its spatial part is related to 
the stress tensor $\sigma_{ij}$ as $\sigma_{ij}=-\mathcal{T}_{ij}$
with $i,j=1,2,3$.
The stress tensor is a fundamental observable
related to force acting on a surface.
In field theory, the stress tensor represents 
distortion of fields induced by external charges~\cite{Landau}.
In Maxwell theory, for example, local propagation of a 
Coulomb interaction between charges is 
characterized by the Maxwell stress tensor, which is 
the spatial component of the EMT in this theory,
$\mathcal{T}_{\mu\nu}=
F_{\mu\rho} F^{\rho}_{\hphantom{\rho}\nu} - (1/4)\delta_{\mu\nu} F_{\rho\sigma}F^{\rho\sigma}$,
with the field strength $F_{\mu\nu}$~\cite{Landau}.
The stress tensor in non-Abelian gauge theories including 
Quantum ChromoDynamics (QCD) is
even more important
because this observable characterizes the structure of the
non-Abelian fields with external sources in a gauge invariant
manner.

In Ref.~\cite{Yanagihara:2018qqg},
the stress-tensor distribution in static quark ($Q$)
and an anti-quark ($\bar{Q}$) systems in vacuum 
in SU(3) Yang-Mills (YM) theory
has been numerically measured on the basis of lattice gauge simulation.
In this study, by utilizing the EMT operator defined on the
lattice~\cite{Suzuki:2013gza} through the 
YM gradient flow~\cite{Luscher:2010iy},
we have shown the local structure of the flux tube
in a gauge invariant way.
We have also quantitatively revealed 
the transverse structure of the stress tensor 
distribution on the mid-plane between $Q\bar{Q}$ by taking 
the continuum limit. 
By employing the Abelian-Higgs model 
as a phenomenological model of QCD~\cite{Yanagihara:2019foh}, 
we have also studied the structure of 
EMT distribution around $Q\bar{Q}$ and compared it with
the numerical results in Ref.~\cite{Yanagihara:2018qqg}.

In this proceedings, we extend our previous study to the analysis
of the stress distribution
in the quark--anti-quark and single quark systems at nonzero temperature.

\section{Energy-Momentum Tensor around Static Quark and Anti-Quark}
\label{sec:emt}
From the EMT, ${\cal T}_{\mu \nu}(x)$ ($\mu,\nu=1,2,3,4$)
in the Euclidean space,
the local energy density and  the stress tensor are respectively given by 
$\varepsilon (x) = -  \mathcal{T}_{44}(x),\,
\sigma_{ij} (x) = - \mathcal{T}_{ij}(x)\ (i,j=1,2,3)$.
The force per unit area $\mathcal{F}_i$
acting on a surface with the normal vector $n_i$
is given by  
$\mathcal{F}_i =  \sigma_{ij}n_j =  - {\cal T}_{ij} n_j$ \cite{Landau}.
The principal axes of stress tensor is obtained after solving 
the eigenvalue equations 
${\cal T}_{ij}n_j^{(k)}=\lambda_k n_i^{(k)} \, (k=1,2,3)$.
Here $n_i^{(k)}$ are the principal axes and 
the strengths of the force per unit area along $n_i^{(k)}$
are given by the absolute values of the eigenvalues $\lambda_k $.
The force acting on a test charge is
obtained by the surface integral ${F}_i =- \int_S {\cal T}_{ij} dS_j$, 
where $S$ is a closed surface surrounding the charge with 
the surface vector $S_j$ oriented  outward from $S$.

Next let us review how we measure the EMT in thermal systems 
with static $Q$ and/or $\bar{Q}$ 
based on the lattice gauge theory at nonzero temperature.
First, at nonzero temperature
static $Q$ and $\bar{Q}$ in Euclidean space are 
represented by the Polyakov loop $\Omega(\vec{x})$ and 
its Hermitian conjugate $\Omega^\dagger(\vec{x})$, respectively.
In the present study
we focus on the singlet $Q\bar{Q}$ 
system~\cite{Maezawa:2011aa,Kaczmarek:2004gv}
and the single $Q$ system.
This system is represented in terms of the correlation function
between the Polyakov loop and its Hermitian conjugate, 
$\mathrm{Tr}[\Omega^\dagger(\vec{x})\Omega(\vec{y})]$.
Since this is a gauge dependent quantity, 
we impose the Coulomb gauge fixing.
In addition to the singlet $Q\bar{Q}$ system, 
we explore the single $Q$,
which is represented by $\mathrm{Tr}\Omega(\vec{x})$,
above the critical temperature $T_c$.

An expectation value of an operator $\mathcal{O}(x)$
in the singlet $Q\bar{Q}$ and the single $Q$ systems
are respectively obtained by
\begin{align}
 \langle {\cal O}(x)\rangle_{Q\bar{Q}} &=
 \frac{\langle  {\cal O}(x) 
 \mathrm{Tr}[\Omega^\dagger(\vec{y})\Omega(\vec{z})]\rangle}
 {\langle \mathrm{Tr}[\Omega^\dagger(\vec{y})\Omega(\vec{z})] \rangle} 
 -\langle  {\cal O}(x)\rangle,
 \label{eq:<T>_QbarQ}
 \\
 \langle {\cal O}(x)\rangle_Q &=
 \frac{\langle  {\cal O}(x)\mathrm{Tr}\Omega(\vec{y})\rangle}
 {\langle \mathrm{Tr}\Omega(\vec{y}) \rangle}
 -\langle  {\cal O}(x)\rangle.
 \label{eq:<T>_Q} 
\end{align}
Note that Eq.~(\ref{eq:<T>_Q}) is ill-defined
below $T_c$ in pure YM theory
because the center symmetry leads 
to $\langle \mathrm{Tr}\Omega(\vec{x}) \rangle=0$ in the confined phase.
We thus consider the single $Q$ system above $T_c$.

In this study, we consider the EMT operator as the observable $\mathcal{O}$
in Eq.~(\ref{eq:<T>_QbarQ}) and Eq.~(\ref{eq:<T>_Q}).
In order to define the EMT in YM theory, we use the YM gradient flow
\cite{Suzuki:2013gza,Luscher:2010iy}.
The YM gradient flow is defined through the flow equation
\begin{align}
\frac{dA_\mu(t,x)}{dt}
= -g_0^2 \frac{\delta S_{\mathrm{YM}}(t)}{\delta A_\mu (t,x)},
\end{align}
where $t$ denotes the fictitious 5-th dimensional 
coordinate called the flow time~\cite{Luscher:2010iy}, 
and the initial condition of $A_{\mu}(t,x)$ at $t=0$ is given by
the ordinary gauge field $A_\mu(x)$ in the four dimensional
Euclidean space.
The YM action $S_{\mathrm{YM}}(t)$ at $t>0$ 
is composed of $A_{\mu}(t,x)$.
The gradient flow for positive $t$ leads to smearing of the gauge field
within the radius $ \sqrt{2t}$.
Using the flowed field, 
the renormalized EMT operator is given by~\cite{Suzuki:2013gza}
\begin{align}
 {\cal T}^{\rm R}_{\mu\nu}(x)=\lim_{t\rightarrow0}{\cal T}_{\mu\nu}(t,x),
 \quad
 {\cal T}_{\mu\nu}(t,x)=c_1(t)U_{\mu\nu}(t,x)
 +4c_2(t)\delta_{\mu\nu}E(t,x) + \mathcal{O}(t),
 \label{eq:T}
\end{align}
where $E(t,x)=(1/4)G_{\mu\nu}^a(t,x)G_{\mu\nu}^a(t,x)$ and
$U_{\mu\nu}(t,x)=G_{\mu\rho}^a(t,x)G_{\nu\rho}^a(t,x)-\delta_{\mu\nu}E(t,x)$
with the field strength $G_{\mu\nu}^a(t,x)$ composed of the flowed gauge field $A_\mu (t,x)$.
We use the higher-order perturbative coefficients for $c_1(t)$ and
$c_2(t)$ obtained in Refs.~\cite{Harlander:2018zpi,Iritani:2018idk}.
The validity and usefulness of this EMT operator
have been confirmed via the study on thermodynamic quantities in SU(3) YM 
theory~\cite{Iritani:2018idk,Kitazawa:2016dsl}.

In lattice simulations we measure
$\langle {\cal T}_{\mu\nu}(t,x)\rangle_{Q\bar{Q}}^{\rm lat}$ 
and $\langle {\cal T}_{\mu\nu}(t,x)\rangle_Q^{\rm lat}$ 
at finite $t$ and $a$.
In order to avoid the discritization effect and
the over-smearing of the gradient flow~\cite{Kitazawa:2016dsl},
one has to choose an  appropriate window of $t$ satisfying the 
condition  ${a}/{2} \lesssim  \rho  \lesssim L$, where
$\rho \equiv \sqrt{{2t}}$ is the flow radius and 
$L$ is the minimal distance between the EMT operator and the Polyakov loop.

Finally we should perform 
an extrapolation to $(t,a)=(0,0)$ in order to obtain the renormalized
EMT operator. In the present study, however, we discuss
preliminary results with fixed $t$ and $a$.

\begin{table*}
 \centering
\begin{tabular}{|cccccc|}
 \hline \hline
 $\beta$ & $a~[\mathrm{fm}]$ & $N_s$ & $N_\tau$ & $N_{\rm conf}$ & $T/T_c$ 
 \\
 \hline
 6.600 & 0.0384 & 48 & 12 & 320 & 1.44 \\
 7.166 & 0.0185 & 48 & 12 & 300 & 2.97 \\
 \hline \hline
 \end{tabular}
 \caption{
 Simulation parameters on the lattice.
 $N_s\,(N_\tau)$ denotes the spatial (temporal) lattice size.
 }
 \label{table:param}
\end{table*}

\section{Setup}
We have performed the numerical simulations 
in SU(3) YM theory on the four-dimensional Euclidean lattice 
with the Wilson gauge action
and the periodic boundary conditions for two temperatures $T$.
The simulation parameters for each $T$ are summarized in Table~\ref{table:param}.
In the measurement of the Polyakov loop $\Omega(\vec{x})$,
we adopt the standard multi-hit procedure 
by replacing every temporal links 
by its thermal average with the neighboring links
for the noise reduction~\cite{Parisi:1983hm}.


\begin{figure}[t]
 \centering
 \includegraphics[width=0.49\textwidth,clip]{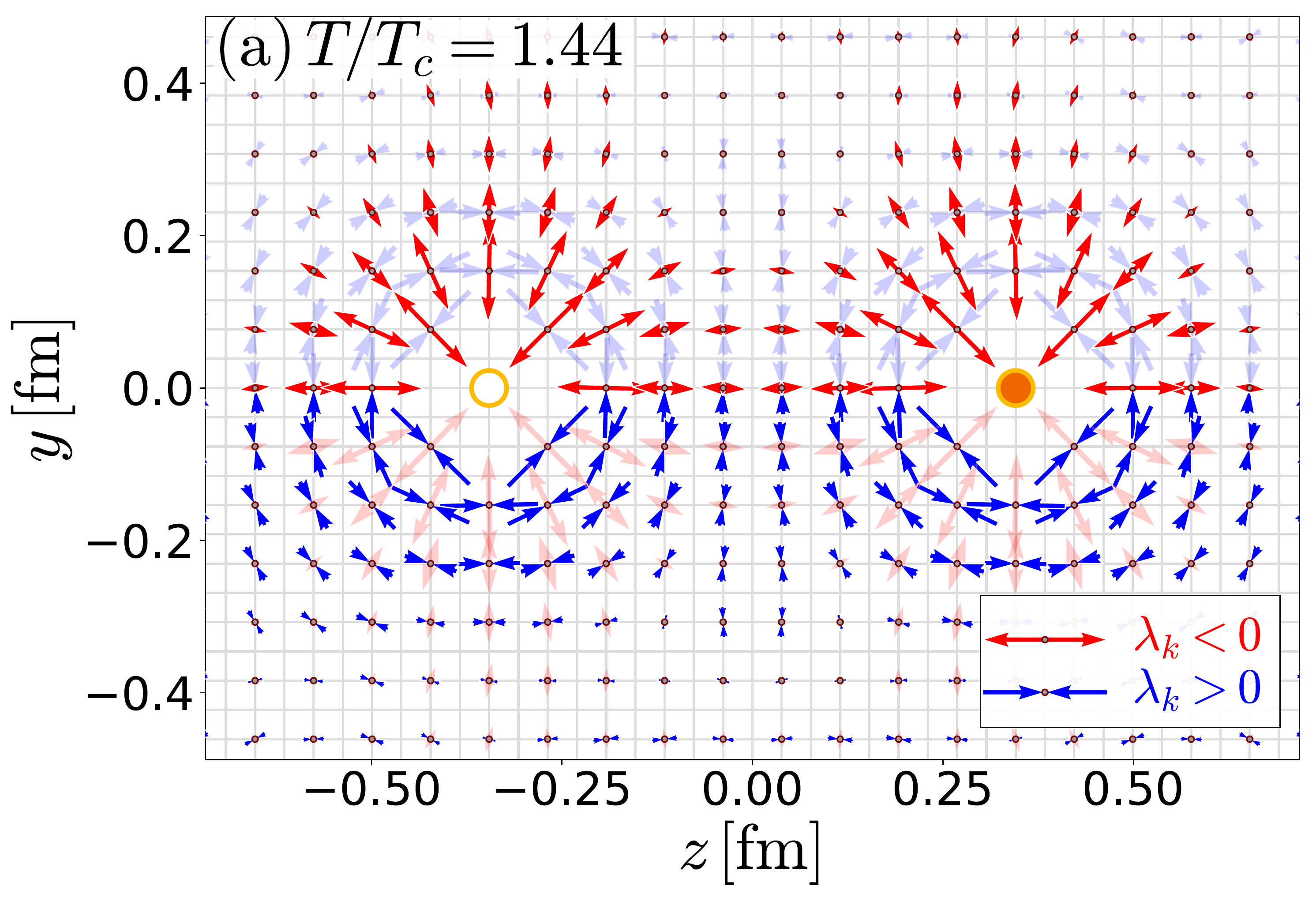}
 \includegraphics[width=0.49\textwidth,clip]{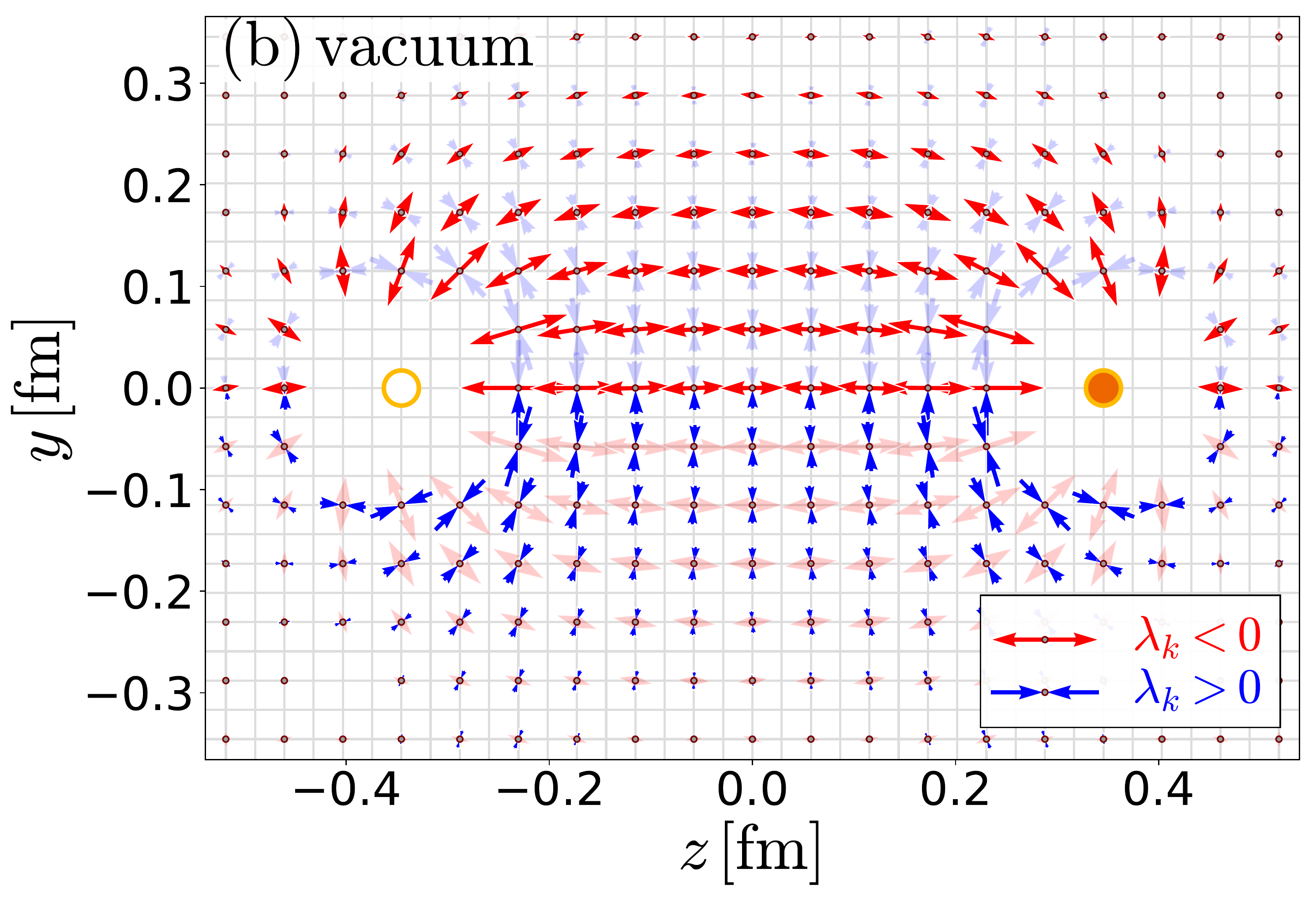}
 \caption{(a) Distribution of the principal axes of ${\cal T}_{ij}$ 
 for a singlet $Q\bar{Q}$ system separated by $R=0.69$ fm 
 in SU(3) Yang-Mills theory 
 with $a=0.038$~fm and $t/a^2=2.0$ at $T/T_c=1.44$.
 (b) Distribution of the principal axes of ${\cal T}_{ij}$
 in vacuum, where $a=0.029$ fm and $t/a^2=2.0$~\cite{Yanagihara:2018qqg}.
 In both panels, the red (blue) arrows in the upper (lower) half plane are 
 highlighted. 
 Note that the lengths of arrrows are suitably scaled.}
 \label{fig:stress-distribution}
\end{figure}

\section{Stress Distribution on the Plane including Two Sources}
In this section, we consider the stress distribution 
in the singlet $Q\bar{Q}$ system, focusing first on the plane including
two sources.
Shown in Fig.~\ref{fig:stress-distribution}~(a) is 
the two eigenvectors of the local stress tensor at $T/T_c=1.44$
around the two sources separated by $R=0.69\ {\rm fm}$
obtained on the lattice with $a=0.038\ {\rm fm}$ 
with fixed $t/{a}^2 = 2.0$.
The other eigenvector is perpendicular to this plane.
The eigenvector with negative (positive) eigenvalue 
is denoted by the red outward (blue inward) arrow
with its length proportional to $\sqrt{|\lambda_k |}$:
\begin{eqnarray}
 \red{\leftarrow} \hspace{-0.05cm} {\tiny{\circ}} \hspace{-0.05cm}  \red{\rightarrow} : \lambda_k<0, 
  \qquad  \blue{\rightarrow} \hspace{-0.05cm}  {\tiny{\circ}}  \hspace{-0.05cm}  \blue{\leftarrow}: \lambda_k>0.
\end{eqnarray}
Neighbouring volume elements are pulling (pushing) with each other
along the direction of red (blue) arrow.   
The arrows in the spatial regions near $Q$ and $\bar{Q}$
, which would suffer from the over-smearing of the gradient flow
due to the overlap between the source charge, are excluded.
In Fig.~\ref{fig:stress-distribution}~(b),
we show the stress distribution around $Q\bar{Q}$ in vacuum
with the same $Q\bar{Q}$ distance $R=0.69\ {\rm fm}$
obtained in Ref.~\cite{Yanagihara:2018qqg} as a comparison.
Fig.~\ref{fig:stress-distribution}~(b) clearly reveals 
the formation of the flux tube in terms of the stress tensor
in a gauge invariant manner;
the region where the strong stress acts concentrates
around the one-dimensional tube structure between $Q\bar{Q}$. 
On the other hand, 
Fig.~\ref{fig:stress-distribution}~(a) shows
that the flux tube, which is formed in vacuum, 
is dissociated at $T/T_c=1.44$ due to medium effects,
and the stress distribution around each source behave alomost independently.

\begin{figure*}[t]
 \centering
 \includegraphics[width=0.49\textwidth,clip]{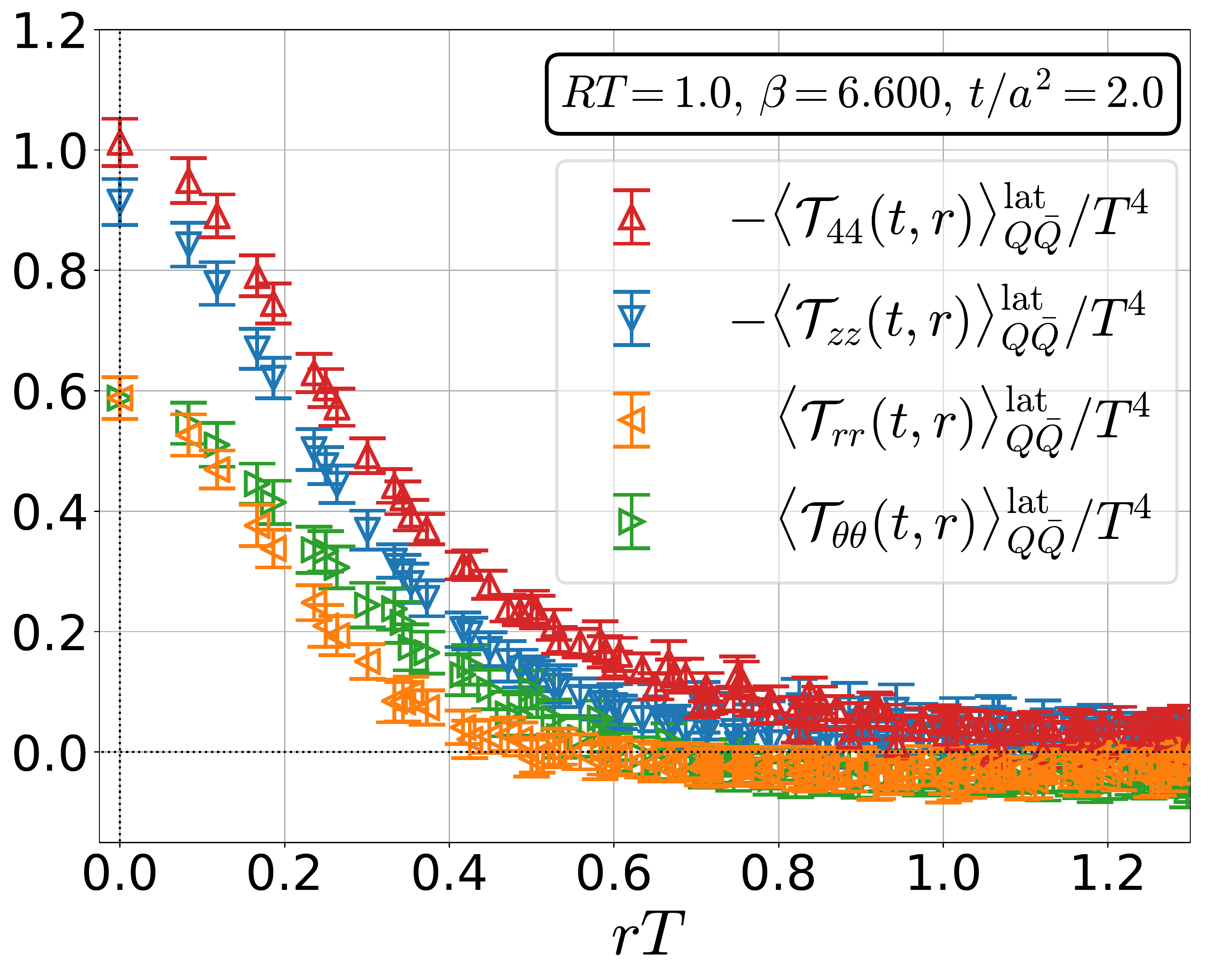}
 \includegraphics[width=0.49\textwidth,clip]{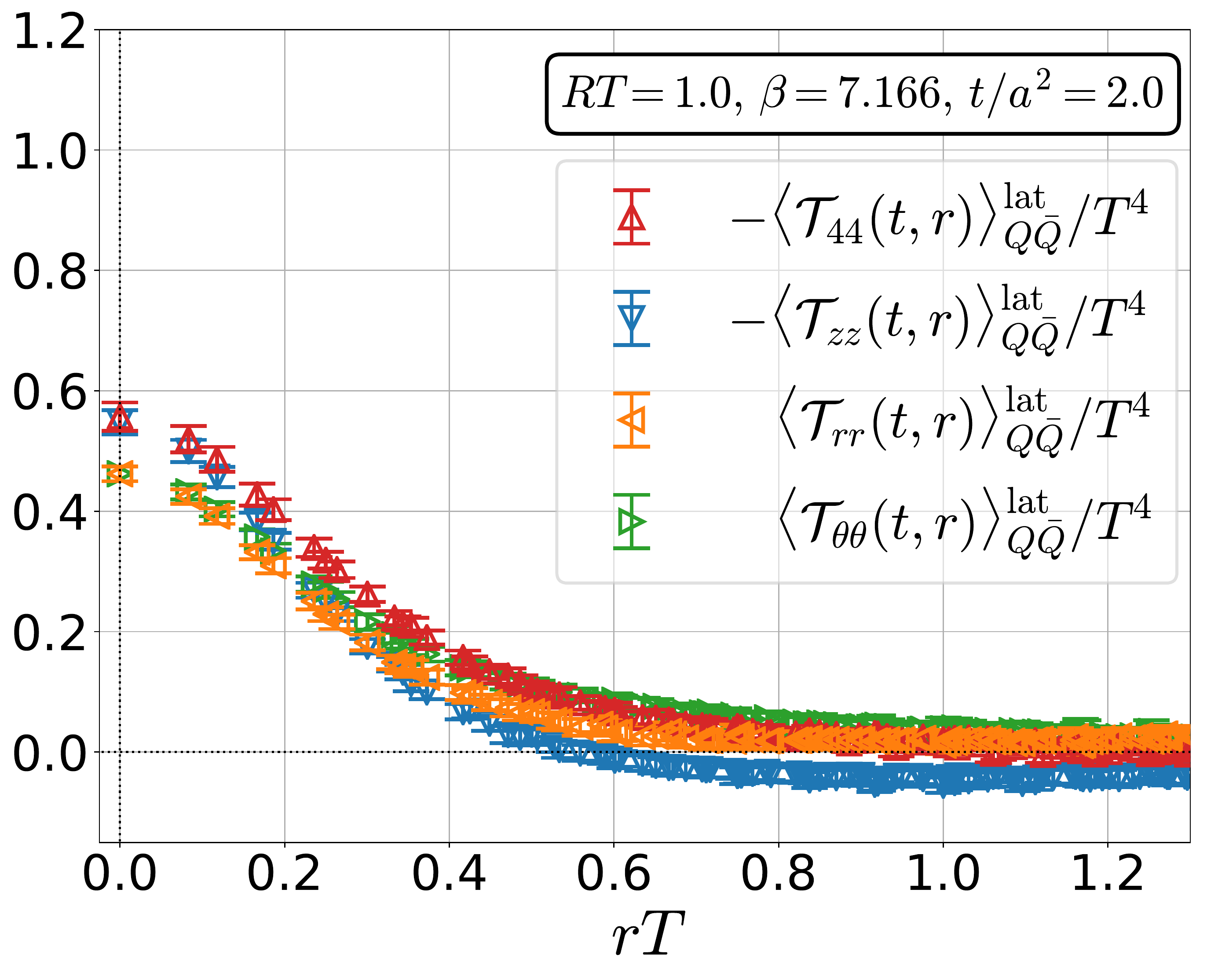}
 \caption{
 EMT distribution on the mid-plane
 $ -\langle{\cal T}_{44}(t,r) \rangle^\mathrm{lat}_{Q\bar{Q}} $ and
 $ -\langle{\cal T}_{cc}(t,r) \rangle^\mathrm{lat}_{Q\bar{Q}} $
 in the cylindrical coordinate system 
 at $T/T_c=1.44\,(\mathrm{left}),2.97\,(\mathrm{right})$ 
 with $RT=1.0$.
 Note that the lattice spacing and the flow time are fixed.
 }
 \label{fig:mid}
\end{figure*}

\section{Stress Distribution on the Mid-Plane between Two Sources}

Next we focus on the mid-plane between $Q$ and $\bar{Q}$.
We use the cylindrical coordinate system $c=(r,\theta, z)$ with
$r=\sqrt{x^2+y^2}$ and $0 \le \theta < 2 \pi$. 
On the mid-plane, because of the cylindrical symmetry and
the parity symmetry with regard to $z$ axis, 
the stress tensor is diagonalized as 
$\langle {\cal T}_{c c'}(t,x) \rangle_{Q\bar{Q}}^{\rm lat} =
{\rm diag} ( \langle{\cal T}_{rr} (t,r)\rangle_{Q\bar{Q}}^{\rm lat},
\langle {\cal T}_{\theta\theta} (t,r)\rangle_{Q\bar{Q}}^{\rm lat},
\langle {\cal T}_{zz} (t,r)\rangle_{Q\bar{Q}}^{\rm lat})$.

In Fig.~\ref{fig:mid}, we show the $r$ dependence of the resulting EMT
at $T/T_c=1.44$ and $2.97$ with $RT=1.0$ in the normalization
$\langle {\cal T}_{c c'}(t,x) \rangle_{Q\bar{Q}}^{\rm lat}/T^4$.
The $Q\bar{Q}$ distance of the left panel in the physical units 
is $R=0.69$~fm.
Note that we fix $t/a^2=2.0$ and the lattice spacing $a$ 
in Fig.~\ref{fig:mid}
and the extrapolation to $(a,t)\to(0,0)$ is not taken.
We notice that the thermal expectation value 
$\langle  {\cal T}_{\mu\nu}(t,x)\rangle$ is subtracted in these results
so that 
$\langle {\cal T}_{c c'}(t,r\rightarrow\infty) \rangle_{Q\bar{Q}}^{\rm lat}=0$.

From the figure and the comparison with the results
in Ref.~\cite{Yanagihara:2018qqg}, one finds several noticeable features:

  \begin{enumerate} 
\setlength{\parskip}{0.1cm}
\setlength{\itemsep}{-0.4cm}
\item Compared with the vacuum result obtained in Ref.~\cite{Yanagihara:2018qqg},
  the absolute values of all components in the left panel
  are suppressed in physical units.
  This suppression is interpreted from the dissociation of the flux tube
  observed in Fig.~\ref{fig:stress-distribution}.
\\ 
\item In the left panel of Fig.~\ref{fig:mid},
  there are approximate degeneracies,
$ \langle {\cal T}_{44}(t,r)  \rangle^{\rm lat}_{Q\bar{Q}}  \simeq
\langle {\cal T}_{zz}(t,r) \rangle^{\rm lat}_{Q\bar{Q}} < 0 $
and  
$ \langle {\cal T}_{rr}(t,r)  \rangle^{\rm lat}_{Q\bar{Q}}  \simeq
\langle{\cal T}_{\theta \theta}(t,r)  \rangle^{\rm lat}_{Q\bar{Q}} >0  $,
for a wide range of $r$ at $T/T_c=1.44$. 
Also, one sees a separation between these two degenerated channels.
From this result, we find 
$ \langle {\cal T}_{\mu \mu}(t,r) \rangle^\mathrm{lat}_{Q\bar{Q}} =
\langle {\cal T}_{44}(t,r) +  {\cal T}_{zz}(t,r) +
{\cal T}_{rr}(t,r)  
+  {\cal T}_{\theta \theta}(t,r) \rangle^\mathrm{lat}_{Q\bar{Q}} <  0 $.
These features are also found in the stress distribution 
in vacuum~\cite{Yanagihara:2018qqg}.
\\

\item By comparing both panels in Fig.~\ref{fig:mid},
  one sees that all components tend to be degenerated in the
  normalization $\langle {\cal T}_{c c'}(t,x) \rangle_{Q\bar{Q}}^{\rm lat}/T^4$
  as $T$ becomes larger.
  This tendency is in part attributed to the fact that
  the value of $R=1.0/T$ for $T/T_c=2.97$ in physical units is smaller than 
  that for $T/T_c=1.44$.
  As $R$ becomes smaller,
  the system is dominated by the physics at high-energy scale and
  the behavior of the EMT around $Q\bar{Q}$ approaches the one
  described by the leading order in perturbation theory
  at which all components of the EMT degenerate. 
\end{enumerate}

\begin{figure*}[t]
 \centering
 \includegraphics[width=0.49\textwidth,clip]{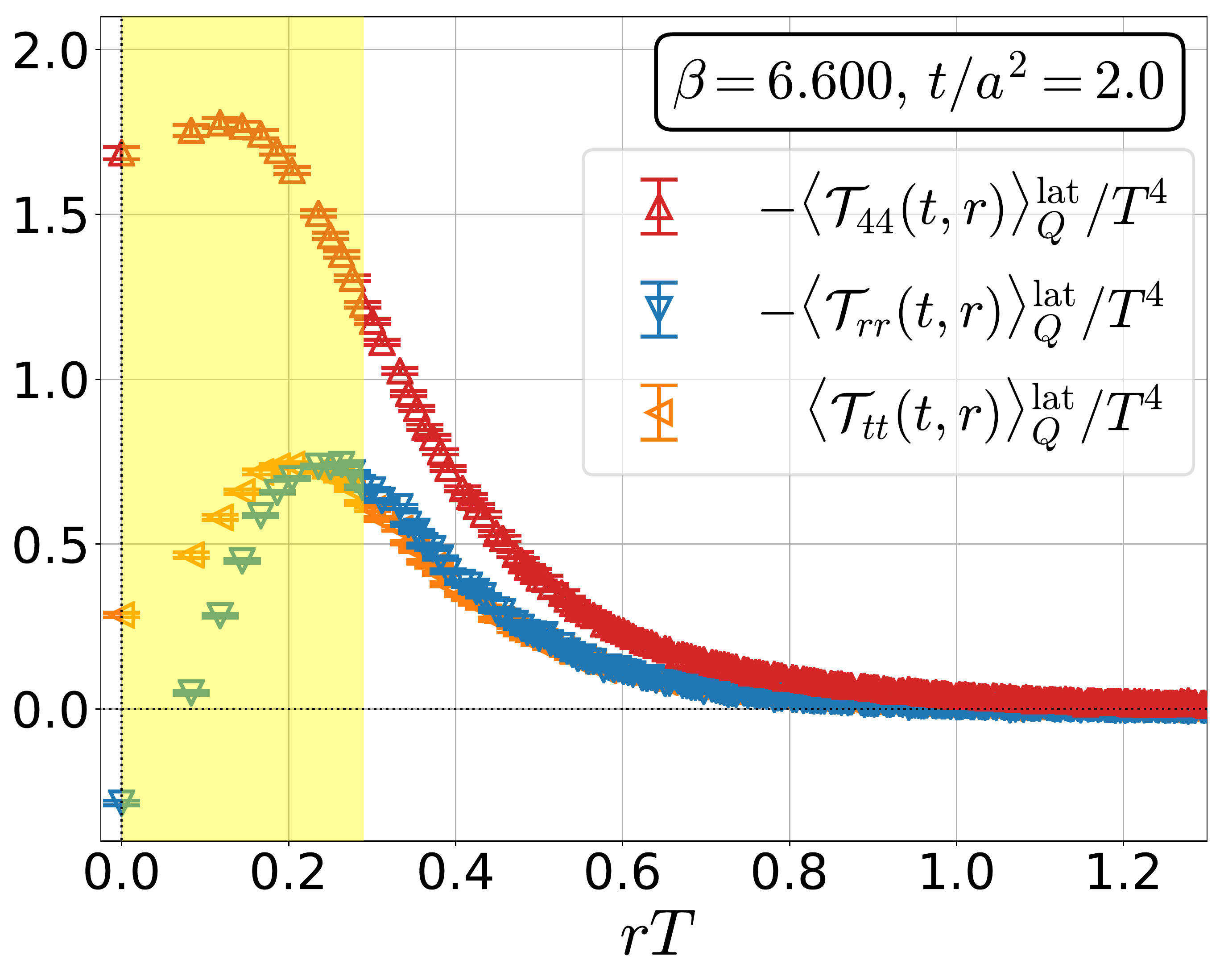}
 \includegraphics[width=0.49\textwidth,clip]{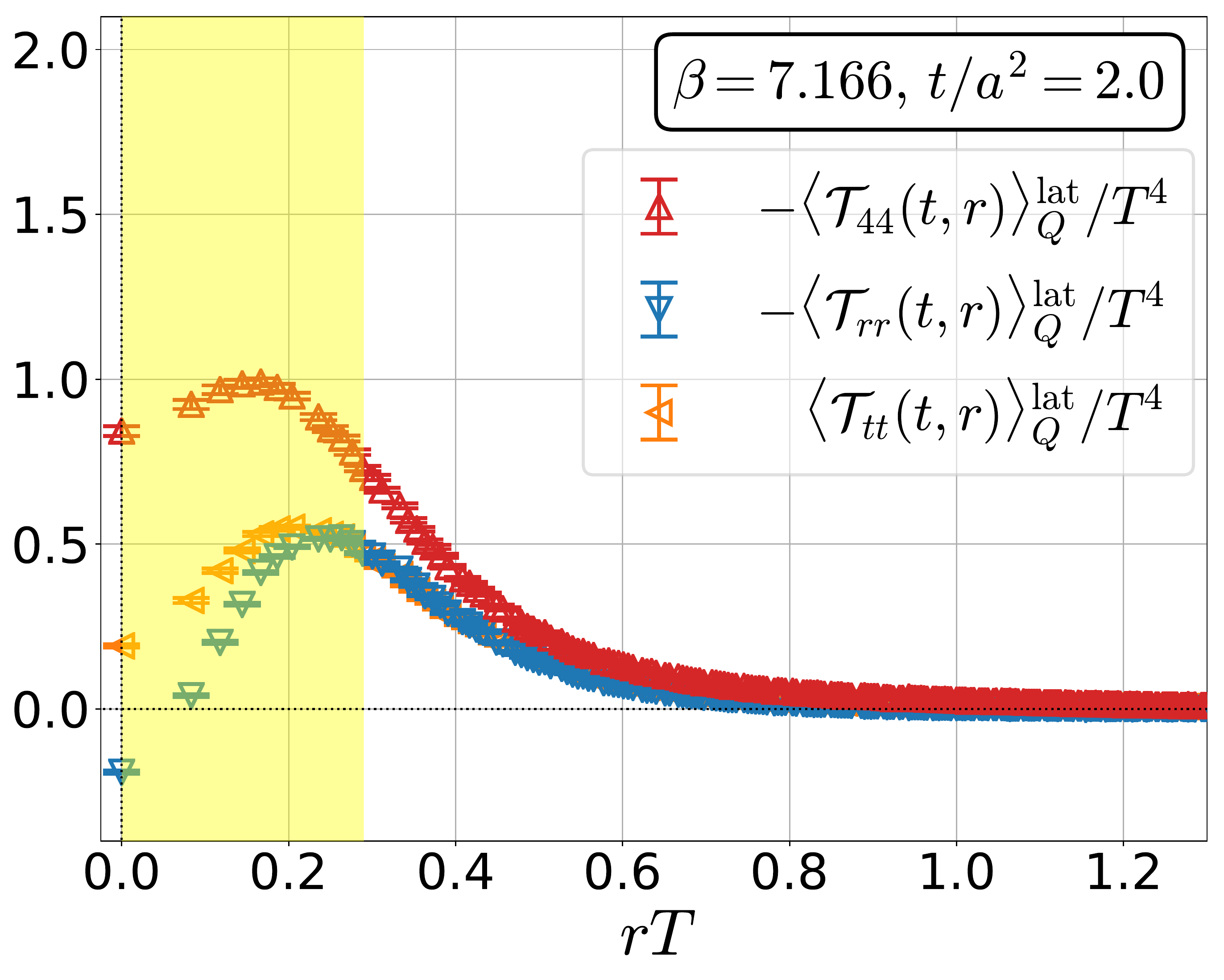}
 \caption{
 EMT distribution around a single static quark
 $ -\langle{\cal T}_{cc}(t,r) \rangle^\mathrm{lat}_{Q} $
 and $ -\langle{\cal T}_{44}(t,r) \rangle^\mathrm{lat}_{Q} $
 in the spherical coordinate system 
 at $T/T_c=1.44\,(\mathrm{left}),2.97\,(\mathrm{right})$.
 Note that the lattice spacing and the flow time are fixed.
 The range of $r$ highlighted by the yellow shades in both figures
 represents the over-smearing region.
 }
 \label{fig:single}
\end{figure*}

\section{Stress Distribution around Single Quark}
Finally we consider the single quark system.
We employ the spherical coordinate system $c=(r,\theta, \varphi)$
with the radial corrdinate is $r=\sqrt{x^2+y^2+z^2}$ and 
the polar and azimuthal angles $\theta$ and $\varphi$.
The spherical symmetry makes  
the stress tensor diagonalized as 
$\langle {\cal T}_{c c'}(t,x) \rangle_{Q}^{\rm lat} =
{\rm diag} ( \langle{\cal T}_{rr} (t,r)\rangle_{Q}^{\rm lat},$
$\langle {\cal T}_{tt} (t,r)\rangle_{Q}^{\rm lat},
\langle {\cal T}_{tt} (t,r)\rangle_{Q}^{\rm lat})$, 
where the transverse components $\langle {\cal T}_{tt} (t,r)\rangle_{Q}^{\rm lat}$
are degenerated owing to the rotational symmetry.
Shown in Fig.~\ref{fig:single} is the profile of each component 
$\langle {\cal T}_{c c'}(t,x) \rangle_{Q}^{\rm lat}$
as a function of $r$ at $T/T_c=1.44$ and $2.97$.
The yellow shades in Fig.~\ref{fig:single}
qualitatively represent the range of $r$ which suffers from
the overlap between the EMT operator and the Polyakov loop;
in this range of $r$, the validity of our analysis is completely lost.

From Fig.~\ref{fig:single}, one finds an approximate degeneracy of
the absolute values of the spatial components $\langle{\cal T}_{rr} (t,r)\rangle_{Q}^{\rm lat}$ and
$\langle{\cal T}_{tt} (t,r)\rangle_{Q}^{\rm lat}$.
The figure also shows that energy densiy $-\langle{\cal T}_{44} (t,r)\rangle_{Q}^{\rm lat}$
has a clear separation from these channels.
One also finds that the magnitudes of all the components become smaller
as $T$ becomes larger.
This behavior is attributed to the EMT at the leading order
in perturbation theory given by $\sim\alpha_s^2(r)/r^4$
with the strong coupling constant $\alpha_s(r)=g^2/(4\pi)$.

From the EMT distribution around a static quark, 
it is expected that many interesting features of the medium can be extracted.
In the large $r$ region, because the color electric field $E$
behaves as $E\sim e^{-m_Dr}$, where $m_D$ is the Debye screening mass, 
all the components of EMT damp exponentially $e^{-2m_Dr}$.
On the other hand, in the small $r$ region
components of EMT behave as $\sim\alpha^2_s(r)/r^4$.
From these behaviors of EMT, one can study the values of $m_D$
and $\alpha_s(r)$.
A numerical confirmation of the mechanical conservation law in the
spherical coordinates, $\partial_r(r^2\mathcal{T}_{rr})=r\mathcal{T}_{tt}$,
is another interesting subject.

\section{Summary and Outlook}
In this proceedings, 
we have explored the spatial distribution of EMT at nonzero temperature 
in the single quark system as well as 
in the singlet $Q\bar{Q}$ system in SU(3) lattice gauge theory. 
The YM gradient flow plays a crucial role to perform these analyses on the lattice.
The dissociation of the flux-tube structure in the $Q\bar{Q}$ system
at high temperature in the deconfined phase
is observed from the stress distribution.

Although we showed the numerical results with fixed $a$ and $t$
throughout this study,
in order to investigate the stress distribution in the continuum limit
one has to take the double extrapolation $(a,t)\to(0,0)$.
Using the double-extrapolated results, we plan to analyze
the static quark systems at nonzero temperature more quantitatively.
In particular, the analysis of 
the Debye screening mass $m_D$ and the strong coupling $\alpha_s(r)$
from the EMT distribution around a single quark is an interesting
future study.
There are also a lot of interesting applications 
of this study, such as 
the generalization to full QCD with the QCD flow
equation~\cite{Makino:2014taa}
and the analyses of the $QQ$ system and the $QQQ$ system
at zero and nonzero temperatures.

\section*{Acknowledgement}

The numerical simulation was carried out on OCTOPUS
at the Cybermedia Center, Osaka University and Reedbush-U
at Information Technology Center, The University of Tokyo.
This work was supported by JSPS Grant-in-Aid for Scientific Researches, 
17K05442, 18H03712, 18H05236, 18K03646, 19H05598.

\end{document}